\documentstyle[prb,aps,epsf]{revtex}
\newcommand{\bphi}{\vec\phi}
\newcommand{\bS}{\vec{S}}
\newcommand{\bl}{\vec{l}}

\begin{document}
%\draft

\twocolumn[\hsize\textwidth\columnwidth\hsize\csname@twocolumnfalse%
\endcsname

\title{Commensurate and incommensurate correlations
       in Haldane gap antiferromagnets}

\author{G\'abor F\'ath and Andr\'as S\"ut\H o}
\address{Research Institute for Solid State Physics, P.O. Box 49,
         H-1525 Budapest, Hungary}

\date{\today}

\maketitle

\begin{abstract}
We analyze the onset of incommensurabilities around the
VBS point of the $S=1$ bilinear-biquadratic model. We propose a simple
effective field theory which is capable of reproducing all known properties
of the commensurate-incommensurate transition at the disorder point $\theta_{\rm vbs}$.
Moreover, the theory predicts another special point $\theta_{\rm disp}$, distinct
from the VBS point, where the Haldane gap behaves singularly. The ground state
energy density is an analytic function of the model parameters everywhere, thus we
do not have phase transitions in the conventional sense.
\end{abstract}

\vskip 0.3 truein
]

\section{Introduction}

The $S=1$ bilinear-biquadratic chain 
\begin{equation}
    H =\sum_{i} \left[\cos\theta\,
         {\rm\bf S}_i {\rm\bf S}_{i+1}
         + \sin\theta\, ({\rm\bf S}_i {\rm\bf S}_{i+1})^2\right]
   \label{spin-1}
\end{equation}
is one of the prototype models for the physics of Haldane 
gap\cite{Hal} antiferromagnets. Its zero temperature phase diagram has been
the subject of intensive studies in recent years.\cite{Aff}$^-$\cite{Sch-Mut}
By now it is well-established that the energy gap (Haldane gap) persists in a 
wide range $-\pi/4<\theta<\pi/4$ around the conventional Heisenberg 
point $\theta=0$. The model is gapless at the special points\cite{TakBabSuth}
$\theta=\pm\pi/4$ beyond
which other phases with qualitatively different physical properties 
appear. Although the gap and the hidden (string) order\cite{denNijs}
characterizing the Haldane-gap systems persist for the whole Haldane phase 
$-\pi/4<\theta<\pi/4$, 
one can divide this interval into (at least) two, somewhat different
subphases.\cite{Bur-Xia-Ger,Sch-Jol-Gar}
These subphases are separated by the so called valence-bond-solid
(VBS) point\cite{AKLT} $\theta_{\rm vbs}=\tan^{-1}{1/3}\approx 0.1024\pi$
where the ground state properties
can be obtained exactly. The two subphases differ in the form of the long
distance asymptotics of the two-point correlation function 
$\langle S^z_i S^z_{i+n} \rangle$. In the 
"commensurate" Haldane phase (C-phase) for $-\pi/4<\theta<\theta_{\rm vbs}$ 
the leading behavior is expected to be
\begin{equation}
   \langle S^z_i S^z_{i+n} \rangle \sim (-1)^n\frac{e^{-n/\xi}}{n^{1/2}} 
   \qquad \mbox{for $n\to\infty$},
   \label{G_c}
\end{equation}
while in the "incommensurate" Haldane phase (IC-phase) for
$\theta_{\rm vbs}<\theta<\pi/4$ this was predicted to take the form
\begin{equation}
   \langle S^z_i S^z_{i+n} \rangle
                \sim \frac{e^{-n/\xi}}{n^{1/2}} \cos (q n +\phi)
   		\qquad \mbox{for $n\to\infty$}
   \label{G_ic}
\end{equation}
where $q=q(\theta)\in (\pi,2\pi/3)$ is a $\theta$-dependent
incommensurate wavenumber, 
$\phi$ is a phase shift, and $\xi=\xi(\theta)$ is the correlation length.
At the "C-IC transition point" (also called the "disorder point") $\theta_{\rm cic}=
\theta_{\rm vbs}$ the correlation function is known 
rigorously and it is purely exponential without any algebraic prefactor
\begin{equation}
   \langle S^z_i S^z_{i+n} \rangle =
   \frac{4}{3} (-)^n e^{-n/\xi}- \frac{2}{3}\delta_{n0},
   \label{G_VBS}
\end{equation}
with $\xi(\theta_{\rm vbs})=1/\ln 3\approx 0.9102$. Correlation functions
of some other operators can also be studied rigorously at the VBS point.\cite{Sch-Mut}
In particular, $\langle (S^z)^2_i (S^z)^2_{i+n} \rangle$ has a similar purely
exponential decay with $\xi=1/\ln 3$ at $\theta=\theta_{\rm vbs}$.

The commensurate-incommensurate (C-IC) transition of the spin-1 chain
bears much similarity to C-IC transitions found in
other models, e.g., in the anisotropic 2D Ising model on the triangular
lattice at finite temperature, where it can be analyzed rigorously
using the exact solution.\cite{Stephenson} Other, higher dimensional, not
exactly solvable models with disorder points were investigated
using an RPA approach to the susceptibility in Ref.\ \ref{Gar-Mai}.
In general, C-IC transitions can be divided into two categories (two kinds).
In this classification scheme the transition
at the VBS point of the $S=1$ chain is a C-IC transition of the {\em first} kind,
with the property that the incommensurate wavenumber $q$ in the IC regime is 
parameter dependent. For C-IC transitions of the first kind the correlation
length is predicted to behave on the C and IC sides of the disorder
point $\theta_{\rm cic}$ as
\begin{equation}
   \left. \frac{d\xi}{d\theta} \right|_{\rm C}=-\infty, \qquad
   \left. \frac{d\xi}{d\theta} \right|_{\rm IC}={\rm finite}
   \label{xi_limit}
\end{equation}
with $\xi(\theta_{\rm cic})\ne 0$ at the transition point.
The characteristic wavenumber is expected to vary on the IC side as
\begin{equation}
   q(\theta) - q(\theta_{\rm cic}) \sim |\theta-\theta_{\rm cic}|^{1/2}.
   \label{q_limit}
\end{equation}
Moreover, the RPA theory\cite{Gar-Mai} also predicts the change of the asymptotic form
of the correlation functions at the disorder point: the algebraic prefactor
is $n^{-(D-1)/2}$ in $D$ dimensions except at the disorder point where
$D\to D^\prime$, $D^\prime<D$ should be taken, reflecting a "dimensional
reduction". In our case $D=1+1=2$, $D^\prime=1$ as is seen in Eqs.\ 
(\ref{G_c}-\ref{G_VBS}). These general features of the C-IC transitions of the first kind
have been tested numerically for the spin-1 chain, and except for some observed
deviation from the predicted form in Eq.\ (\ref{G_ic}) slightly above
$\theta_{\rm vbs}$, all were
justified.\cite{Sch-Jol-Gar} Note that around the VBS point finite size corrections
are very small, thus the numerical results (exact diagonalization and DMRG) are
extremely precise.

\begin{figure}[hbt]
\epsfxsize=\columnwidth\epsfbox{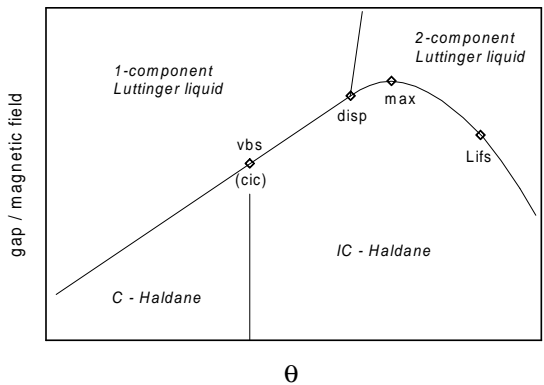} 
\caption{Schematic phase diagram showing the neighborhood of the C-IC
transition with the special points. The critical
magnetic field, where the Haldane gap collapses, equals the value of the
energy gap.} 
\label{fig:specpoints} 
\end{figure}

Within the IC subphase some other special points can be defined. The first one
is $\theta_{\rm disp}\approx 0.1210\pi$, where the second derivative of
the magnon dispersion at $k=\pi$ vanishes, and the dispersion becomes quartic.
When the chain is subject to
a uniform magnetic field the magnetization-vs-field curve has an
anomalous, non-square-root-like singularity at this
point.\cite{Oku-Hie-Aku} For $\theta>\theta_{\rm disp}$, the second
derivative at $k=\pi$ is negative and
the gap takes its minimum value for a momentum different from
$k=\pi$.\cite{Gol-Jol-Sor} In this region
a strong enough field causes the Haldane gap to collapse into a
{\em two}-component Luttinger liquid (LL) phase instead of a
conventional one-component LL phase.\cite{Fat-Lit} The next special point is
$\theta_{\rm max}\approx 0.123\pi$ where the Haldane gap takes its maximum value.
Note that this point has less physical relevance, since it strongly depends
on the actual definition of the Hamiltonian; a $\theta$-dependent rescaling
of the model can easily change its location. Finally one can define
the Lifsitz point $\theta_{\rm Lifs}\approx 0.1314\pi$, where the
incommensurability manifests itself in the structure factor. This is different
from the disordered point in massive models due to the finite linewidth
of the peaks.\cite{Sch-Jol-Gar}
Figure \ref{fig:specpoints} shows these special points.

%%%%%%%%%%%%%%%%%%%
%%%%%%%%%%%%%%%%%%%

In both the C and IC Haldane phases the energy spectrum consists of a discrete
triplet branch of "magnons" separated by finite gaps from the singlet
ground state, and also from a continuum of higher lying "multi-magnon"
excitations in a wide momentum range. This one-magnon branch
is clearly discernible around the edges of the Brillouin zone $k=\pm\pi$.
However, it merges into the continuum and vanishes due to magnon
scattering processes in an extended range around $k=0$.
The energy spectrum, especially in the Heisenberg point and
the VBS point was studied numerically by many
authors.\cite{Sch-Mut,Whi-Hus,Fat-Sol-CM,Aro-Aue-Hal} All concluded that the
magnon-magnon interactions are rather weak, and bound states do not play
a role at low energies. Many properties of the system, from ground-state
correlation functions\cite{Whi-Hus,Sor-Aff} to the onset of magnetization
in uniform fields\cite{Oku-Hie-Aku,Sak-Tak,Aff-Sor} can be extremely well approximated
using a massive relativistic free boson theory. Such an approximate theory
can be derived directly from the non-linear $\sigma$-model (NL$\sigma$M)
description of the spin chain at the Heisenberg point,\cite{NLSM} or from
the Majorana fermion representation of the integrable Takhtajan-Babujian
model in the vicinity of $\theta=-\pi/4$.\cite{Tsvelik} However, all these
microscopic theories utilize the {\sl a priori} assumption that the
important low-energy
fluctuations are at momenta $k=0$ and $\pi$, and thus they cannot account
for the C-IC transition, nor can they give any reliable description of the
IC regime.

Although we do not have a rigorous microscopic justification
all results we possess are consistent with the assumption that the
elementary excitations are essentially free bosons in the IC regime, too.
Thus the aim of the present paper is to extend the free boson description to
the whole Haldane phase, and give a general theory which is capable of accounting for
the C-IC transition in simple terms. In the lack of a detailed microscopic
formulation, however, this theory, at present, is only phenomenological.

%%%%%%%%%%%%%%%%%%%%%%%%%%%%%%%%%%%%%%%%%%%%%%%%%%%
\section{Effective theory - continuum version}

A continuum field theory to describe Heisenberg antiferromagnetic
chains ($\theta=0$)
with integer $S$ was developed by Haldane.\cite{Hal} This is the nonlinear
$\sigma$-model (NL$\sigma$M) -- without topological terms -- which can be
derived in the large $S$ limit, but whose implications are believed to
hold for $S=1$ too. The NL$\sigma$M is defined by the Lagrangian\cite{NLSM}
\begin{equation}
   {\cal L}=\frac{1}{2g}\left[{1\over v}
   (\partial_t\bphi)^2 -v(\partial_x\bphi)^2\right],
   \label{Lagr-NLsigmaM}
\end{equation}
where $\bphi(x,t)$ is a vector field with unit length $\bphi^2=1$,
$v$ is a nonuniversal constant (velocity) setting the energy scale,
and $g=2/S$ is the coupling constant. The
associated Hamiltonian is
\begin{equation}
   {\cal H}=\frac{v}{2}\left[
    g \bl^2 +{1\over g}(\partial_x\bphi)^2\right],
   \label{Ham-NLsigmaM}
\end{equation}
where the momentum canonically conjugate to $\bphi$ turns out to be
\begin{equation}
   \bl(x,t) \equiv {1\over gv} \left[\bphi\times\frac{\partial\bphi}{\partial t}
                   \right](x,t).
   \label{l-term}                
\end{equation}
The spin operator $\bS_n$ can be expressed in terms of $\bphi$
and $\bl$ as
\begin{equation}
   \bS_n(t) = (-1)^n S \bphi(x_n,t) + \delta x\, \bl(x_n,t)           
   \label{S-phi-l}
\end{equation}
where $x_n=n\delta x$, $\delta x$ being the lattice constant
(usually set to unity), and $S=1$ in the present case.
At the mean field level the NL$\sigma$M
can be well approximated by a massive, essentially free vector-boson theory\cite{}
\begin{equation}
   {\cal L}={1\over v}
   (\partial_t\bphi)^2 -v(\partial_x\bphi)^2  - m^2 \bphi^2,   
   \label{relat-Lagr}
\end{equation}
now without any constraint on the $\bphi$ field. 
The boson field $\bphi$ varies
smoothly on the scale of the lattice constant in the commensurate
regime (the pure Heisenberg model, which the theory applies for, is in the C phase)
and thus higher
derivatives, neglected in Eq.\ (\ref{relat-Lagr}), are indeed small.
The Lagrangian gives rise to a relativistic dispersion $\omega(k)=
\sqrt{m^2+v^2 k^2/2m}$ which takes its minimum at $k=0$. [Note the factor
$(-1)^n$ introduced in Eq.\ (\ref{S-phi-l}) shifting all momenta by $\pi$.]

The NL$\sigma$M summarized above gives a good description of the low-energy
behavior of the $S=1$
Heisenberg chain, but is unable to describe the C-IC transition
in the bilinear-biquadratic chain. The key feature missing is
that in the vicinity of our special points
the shape of the magnon dispersion changes drastically, and the minimum at
$k=\pi$ ($k=0$ in the momentum-shifted boson language) splits. Obviously,
such an effect can never be obtained from
a relativistic field theory such as Eq.\ (\ref{relat-Lagr}) and we need to
consider a non-relativistic model.
The simplest continuum Lagrangian with this property is
\begin{eqnarray}
   {\cal L}%&=&{\cal L}(\phi,\partial_t\phi,\partial_x\phi,
           % \partial_x^2\phi)\nonumber\\
           &=& 
   (\partial_t\bphi)^2 -a\, (\partial_x\bphi)^2
   -b\,(\partial_x^2\bphi)^2 - m^2 \bphi^2.
   \label{Lagr}
\end{eqnarray}
At this level 
$a$, $b$, and $m$ are $\theta$-dependent phenomenological parameters.
For stability $b$ is supposed
to be positive, and $a$ is assumed to change sign at the special point
$\theta_{\rm disp}$. [Note that in the relativistic case $b=0$, and 
$a$ is in fact $v^2$, where $v$ is the "speed of light". Formally
taking $a<0$, which we will consider in the sequel, implies an imaginary $v$.
This only means, however, that one should redefine the physical (quasi)particles
to live around the two new minima of the dispersion. In fact, one
should introduce two new particles, now with $v=$real, for the two minima.
This immediately leads to a two-particle (two-band) description which was
analyzed in detail in Ref.\ \ref{Fat-Lit-98} for the massless case
(above the critical magnetic field) and was shown to result a two-component
(two-band) Luttinger liquid there.
Here we shortcut these complications by simply assuming that the two particles
are two chiral components ($k<0$ and $k>0$) of the same boson with $a<0$.]
%The presence of the second spatial derivative in the potential energy reflects the
%fact that the underlying microscopic model is defined on a lattice. This point
%will be elucidated in the next Section.
%It may stem
%from the Taylor expansion of a term like $[\phi(x+\partial x)-\phi(x)]^2 
%/\Delta x^2$ ($\Delta x$ being the lattice constant),
%which replaces $(\partial_x\phi)^2$ of the continuum theory.

Since our phenomenological model is not derived directly from the
microscopic Hamiltonian in Eq.\ (\ref{spin-1}) the connection between 
the boson field and the original spin variables is not known rigorously.
In principle $\bS_n$ can be expanded in powers of the field $\bphi$ and
its derivatives, and any term permitted by the symmetries of the model
can appear. Being guided by the NL$\sigma$M description at the Heisenberg
point, in the following we will assume that the first two
terms (linear and quadratic in $\bphi$) are
\begin{equation}
   \bS_n(t) = g_\phi (-1)^n\bphi(x_n,t) +
              g_l \bl(x_n,t)
              +{\cal O}(|\bphi|^3),           
   \label{Szphi}
\end{equation}
where $g_\phi$ and $g_l$ are unknown constants. Note that the $\bl$ term, defined
by Eq.\ (\ref{l-term}),
is the most relevant 2-boson term which is even under parity and odd under
time reversal as it should be. Although, we cannot exclude the
possibility that up to ${\cal O}(\bphi^2)$ some other terms with higher
derivatives also appear, such extra derivatives,
due to the finite mass gap, would not modify the
long distance asymptotics of $\langle \bS_0 \bS_n \rangle$.
Equation (\ref{Szphi}) shows that in the C phase where spin-spin correlations
are antiferromagnetic $\phi$ varies smoothly, and higher order derivatives in
the Langrangian have minor role. Around the C-CI transition point, and
in the IC-phase, however, this is no longer the case: as $\bS_n$ picks up
incommensuration, $\bphi(x)$ must do so, as well. It no longer varies smoothly,
and higher derivatives in the Lagrangian cannot be neglected.
This is the effect which we try to take into consideration by the $b$ term in Eq.\
(\ref{Lagr}).

The
asymptotic behavior of the correlation function $\langle S^z_i S^z_{i+n}
\rangle$ is determined by the one-boson term; the two-boson term
and in general any multi-boson terms only constitute minute corrections
which decay at least twice as rapidly. Thus, up to the two-boson
term in Eq.\ (\ref{Szphi}) the spin-spin correlation function
has the behavior
\begin{equation}
   \langle S_n^z(t) S_0^z(0)\rangle = g_\phi^2 (-)^{n} G_\phi(x_n,t) +
                                g_l^2 G_l(x_n,t),
   \label{corr}
\end{equation}
with
\begin{eqnarray}
   G_\phi(x,t) &\equiv& \langle \phi^z(x,t)\phi^z(0,0) \rangle , \\
   G_l(x,t)&\equiv& \langle l^z(x,t) l^z(0,0) \rangle .
\end{eqnarray}
These are the quantities we will now calculate.

The Euler-Lagrange equation associated with our Lagrangian gives a
generalized Klein-Gordon equation for each component $\alpha=x,y,z$ of $\bphi$
\begin{equation}
    \partial_t^2\phi^\alpha - a\, \partial_x^2\phi^\alpha
   +b\, \partial_x^4\phi^\alpha+ m^2\,\phi^\alpha =0,
\end{equation}
whose Green's function is
\begin{equation}
   G_\phi(\omega,k) = \frac{1}{\omega^2-a k^2-b k^4-m^2 +i\varepsilon}.
\end{equation}
This defines the non-relativistic dispersion
\begin{equation}
   \omega(k)=\sqrt{m^2+a k^2+b k^4},
   \label{disp}
\end{equation}
which reduces to the relativistic dispersion of Ref.\ \ref{Sor-Aff}
when $b=0$. 

The generalized theory
can be quantized identically to the relativistic Klein-Gordon theory.\cite{Kak}
The field $\phi^\alpha$, $\alpha=x,y,z$, has the following mode expansion
\begin{equation}
   \phi^\alpha(x,t)=\int \frac{dk}{\sqrt{4\pi\omega(k)}}
           \left[d^\alpha_k e^{i{\bf K\cdot X}}+
                 d^{\alpha\dag}_k e^{-i{\bf K\cdot X}}\right]
   \label{mode}              
\end{equation}
where ${\bf K\cdot X}=\omega t-k x$, and the normalization is
$[d^\beta_k,d^{\alpha\dag}_{k^\prime} ]=\delta_{\beta\alpha}\delta(k-k')$.
Using the mode expansion the equal time expressions $G_\phi(x)$ and $G_l(x)$
can be easily reduced to Fourier transforms\cite{Sor-Aff}
\begin{eqnarray}
   G_\phi(x) &=&
   \int \frac{dk}{4\pi} \frac{e^{ikx}}{\omega(k)}, \nonumber\\
   \label{phiphi} \\
   G_l(x) &=& {1\over 2} \int \frac{dk'}{4\pi} \omega(k')e^{ik'x}
       \int \frac{dk}{4\pi} \frac{e^{ikx}}{\omega(k)} -{1\over 2}\delta^2(x).
       \nonumber
\end{eqnarray}
Note that the integral determining $G_\phi$ also appears as a multiplicative
factor in $G_l$.

%%%%%%%%

Since the asymptotic behavior is determined by $G_\phi$, we start our analysis
with this. The evaluation of the Fourier transform starts with locating
the zeros of $\omega(k)$ in the complex plane. The four zeros are given by
\begin{equation}\label{zeros}
   k=\pm
   \left[{1\over 2b}\left(-a\pm\sqrt{a^2-4m^2b}\right)\right]^{1\over 2}
\end{equation}
and depend on the single parameter $\theta$, via $a$, $b$ and $m$.
Since our phenomenological model is not derived directly from the microscopic
one, we have to make several assumptions about the $\theta$-dependence of
$a$, $b$ and $m$. We will assume that this dependence is smooth 
(analytical), $m(\theta)$ and $b(\theta)$ are nonnegative
whereas $a$ decreases with increasing $\theta$
and changes sign at $\theta_{\rm disp}$. We introduce the discriminant
\begin{equation}
   D(\theta)=a^2-4m^2b,
   \label{discriminant}
\end{equation}
which is hence an analytic function of $\theta$, and we suppose that it
is positive for
$\theta<\theta_{\rm vbs}$, vanishes at the VBS point, and it is negative from
$\theta_{\rm vbs}$ to $\theta=\pi/4$ where it vanishes again. 
As we show below, under the above hypotheses the phenomenological model
provides the expected asymptotic behavior of the correlation function
in all parts of the Haldane phase.\medskip

%%%%%%%%%%%%%%%%%%%
\begin{figure}[hbt]
\epsfxsize=\columnwidth\epsfbox{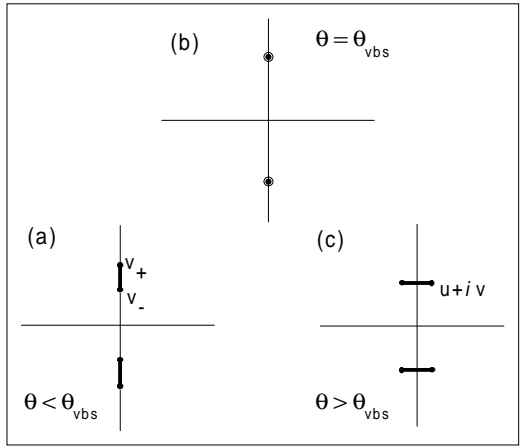} 
\caption{Roots of $\omega(k)$ in the complex $k$ plain. (a) $\theta<
\theta_{\rm vbs}$, (b) $\theta=\theta_{\rm vbs}$, (c) $\theta>
\theta_{\rm vbs}$. Branch cuts are chosen according to the conventions
described in the text.} 
\label{fig:cuts} 
\end{figure}
%%%%%%%%%%%%%%%%%%%

\leftline{\boldmath{$\theta=\theta_{\rm vbs}$}:}

At this point $a>0$ and $D=0$. The expression under the square-root in
$\omega(k)$ is a complete square, thus $\omega$ becomes quadratic in
$k$ and has two purely imaginary zeros, 
$\pm i\sqrt{2m^2/a}=\pm i\sqrt{a/2b}$ [see Fig.\ \ref{fig:cuts}(b)]. 
The contour of integration can be closed in the upper half plane and the
result is
\begin{equation}\label{VBS}
G_\phi(x)={1\over 2\sqrt{2a}}e^{-\sqrt{a\over 2b}|x|}\ .
\end{equation}
Because of the purely exponential decay, this indeed corresponds to 
$\theta=\theta_{\rm vbs}$.\medskip

\leftline{\boldmath{$\theta<\theta_{\rm vbs}$}:}

In this region
$a>0$ and $D>0$. We may suppose $b>0$; the case when $b=0$ can
be obtained by continuity. We get 
four purely imaginary zeros $\pm iv_\pm$ where
\begin{equation}\label{vpm}
v_\pm= 
\left[{1\over 2b}\left(a\pm\sqrt{D}\right)\right]^{1\over 2}\ .
\end{equation}
Now
\begin{equation}
\omega(k)=\sqrt{b}(k^2+v_+^2)^{1\over 2}(k^2+v_-^2)^{1\over 2}
\end{equation}
is single-valued in the complex plane with two cuts, one between  
$-iv_+$ and $-iv_-$ and another one between $iv_-$ and $iv_+$
[see Fig.\ \ref{fig:cuts}(a)]. (We use the convention that $z^{1/2}$
has a cut along the negative real axis.)
The contour of integration can be closed in the upper half-plane and drawn
onto the upper cut, giving
\begin{equation}\label{AF}
G_\phi(x)={1\over 2\pi\sqrt{b}}\int_{v_-}^{v_+}{e^{-t|x|}\over
     \left[(t^2-v_-^2)(v_+^2-t^2)\right]^{1\over 2}} dt \ .
\end{equation}
This function is positive for all $x$, so with the factor $(-1)^n$
introduced in Eq.\ (\ref{Szphi}) we obtain
the expected antiferromagnetic modulation. As $|x|$ goes to
infinity, the main contribution to the integral is coming from
the vicinity of the end point $v_-$, and it is
legitimate\cite{Watsonlemma} to expand
the integrand around this point. For $|x|>|D|^{-1/2}$ this yields
\begin{equation}\label{AFasy}
   G_\phi(x)={e^{-v_-|x|}\over 2[2\pi v_-\sqrt{D}]^{1\over 2}}
\left[|x|^{-{1\over 2}}+O(|x|^{-{3\over 2}})\right]\ .
\end{equation}
Together with Eq.\ (\ref{Szphi}), we find the expected asymptotic form
Eq.\ (\ref{G_c}) of the spin-spin correlation function with a
correlation length $\xi=1/v_-$ which is continuous
at $\theta=\theta_{\rm vbs}$.\medskip

\leftline{\boldmath{$\theta>\theta_{\rm vbs}$}:}

Here D becomes negative and $\omega$ has
four complex zeros $k_j=\pm u\pm iv$ ($j=1,\ldots,4$) with
\begin{equation}\label{uv}
   u=\left[{m\over 2\sqrt{b}}-{a\over 4b}\right]^{1\over 2} 
   \qquad
   v=\left[{m\over 2\sqrt{b}}+{a\over 4b}\right]^{1\over 2} \ .
\end{equation}
If we write $\omega$ in the form
\begin{equation}
\omega(k)=\sqrt{b}\prod_{j=1}^4(k-k_j)^{1\over 2}
\end{equation}
we see that it is single-valued on the complex plane with a cut between
$-u-iv$ and $u-iv$ and another cut between $-u+iv$ and $u+iv$
[see Fig.\ \ref{fig:cuts}(c)]. 
For $k$ real we get back the original positive function.
The integration can be carried out along a contour which starts at $-u+i\infty$,
goes vertically down to $-u+iv$, passes below the upper cut and goes vertically
to $u+i\infty$. Next, we replace the integral below the cut by an integral going
above the cut and in the opposite sense, and this latter by the sum of the
two integrals along the vertical half-lines. These are complex conjugate to
each other, so finally we obtain
\begin{eqnarray}\label{IC}
   G_\phi &&(x)= \\
   &&  {1\over\pi\sqrt{b}}\int_v^\infty{e^{-|x|t}\cos[u|x|-\varphi(t)]\; dt\over
     \sqrt{t^2-v^2}[4u^2+(t-v)^2]^{1\over 4}[4u^2+(t+v)^2]^{1\over 4}},
     \nonumber
\end{eqnarray}
where 
\begin{equation}
\varphi(t)={1\over 2}\left(\arctan{t-v\over 2u}+\arctan{t+v\over 2u}\right).
\end{equation}
Now $G_\phi(x)$ changes sign periodically, and we can identify $\pi-u$
with the wave number $q$ of the incommensurate oscillation.
At the VBS point $u=0$, and as it is shown by Eq.\ (\ref{uv}), the
assumed analyticity of $m$, $a$ and $b$ assures that it has a square-root-type
singularity above the VBS point in accordance with Eq.\ (\ref{q_limit}).
The large-$|x|$ asymptotics of $G_\phi(x)$
can be obtained by Watson's lemma\cite{Watsonlemma} or by a direct expansion,
%\begin{equation}\label{ICasy}
\begin{eqnarray}\label{ICasy}
    G_\phi &&(x)= \\
&& {e^{-v|x|}\cos(u|x|-{1\over 2}\arctan{v\over u}) \over
     (2\pi)^{1\over 2} \left(\frac{-mD}{\sqrt{b}}\right)^{1\over 4}}
     \left[|x|^{-{1\over 2}}+O(|x|^{-{3\over 2}})\right].\nonumber
%\end{equation}
\end{eqnarray}

Formulas (\ref{IC}) and (\ref{ICasy}) apply for any $\theta$ between $\theta_{\rm vbs}$
and $\pi/4$,
including $\theta_{\rm disp}$ which is a symmetry point of the
domain of incommensurate oscillations ($u=v$). As $\theta$ approaches $\pi/4$,
$D$ goes to zero and the zeros of $\omega(k)$ tend to the real axis. Thus,
$v$ goes to zero and the correlation length diverges. This is what we expect
at the boundary of the Haldane phase $\theta=\pi/4$ where the gap disappears.
At this special point the ground state has a tripled periodicity,\cite{FS123,TakBabSuth}
implying $u(\theta=\pi/4)=2\pi/3$.\medskip

For {\em any fixed} $x$, $G_\phi(x)$ depends analytically on $\theta$ inside the whole
Haldane phase. This can be seen from the original form Eq.\ (\ref{phiphi})
of $G_\phi(x)$ by inserting the original, non-factorized
expression Eq.\ (\ref{disp}) for $\omega(k)$. A proof can be found in [\ref{Tit}].
The argument makes use of the continuity of the integrand in $k$ real, its (supposed)
analyticity for any fixed $k$ as a function of $\theta$ in suitable complex domains, and the
uniform convergence of the integral for $\theta$ in any of these domains.
It is interesting to examine another kind of asymptotics, valid in a close
neighborhood of $\theta_{\rm vbs}$, when $|D|/a^2\ll 1$. In this case Eq.\
(\ref{AF}) reduces to the form
\begin{eqnarray}
   G_\phi(x) &\approx& {e^{-\sqrt{a\over 2b}|x|}\over \sqrt{8a}}{1\over \pi}\int_0^\pi
   \cosh\left[\sqrt{D\over 8ab}|x|\sin\alpha\right] d\alpha \nonumber\\
   &\equiv&
   {e^{-\sqrt{a\over 2b}|x|}\over \sqrt{8a}}
   I_0\left(\sqrt{D\over 8ab} |x|\right)  \label{approx}
\end{eqnarray}
where $I_0$ is the zeroth order Bessel function.
We can arrive at the same equation from Eq.\ (\ref{IC}), by changing the contour
of integration (integrating around the upper cut).
Now analyticity at the VBS point is manifest,
because in the expansion of the hyperbolic cosine about zero only the even
powers of $\sqrt{D}$ appear. Equation (\ref{approx})
shows that the crossover to the decay with the
$|x|^{-{1\over2}}$ prefactor sets in at the characteristic distance
$|x|\sim D^{-{1\over 2}}$, which diverges at the VBS point. This explains
the numerical difficulties\cite{Sch-Jol-Gar} verifying the expected asymptotic
behavior very close to the VBS point.

At the VBS point there is an infinite jump in the
derivative of the correlation length, as predicted by Eq.\ (\ref{xi_limit}).
Indeed, Eq.\ (\ref{vpm}) yields  
\begin{equation}
   \left.{d\xi\over d\theta}\right|_{\theta_{\rm vbs}-0}
   \left.\approx {D'\over4\sqrt{2a}m}\right|_
   {\theta_{\rm vbs}}D^{-{1\over 2}}
   \sim -(\theta_{\rm vbs}-\theta)^{-{1\over2}}
\end{equation}
because $D'(\theta_{\rm vbs})<0$. On the other hand, from Eq.\ (\ref{uv})
\begin{equation}
   \left.{d\xi\over d\theta}\right|_{\theta_{\rm vbs}+0}
   =\left.-{1\over2\sqrt{a}}\left({m'\over 2m}+{a'\over 2a}+{m^2\over
   a^2}\right)\right|_{\theta_{\rm vbs}},
\end{equation}
which is finite. The singularity of the correlation length at 
$\theta_{\rm vbs}$ is in no contradiction with the analyticity of $G(x)$
at a {\em fixed} $x$. Indeed, the divergence of the derivative of $\xi$ was
extracted from the single-exponential asymptotic form Eq.\ (\ref{AFasy}) which,
again, is valid only for $|x|>(v_+-v_-)^{-1}\sim D^{-{1\over 2}}$.

To see the role of the two-boson term $G_l$ in the correlation functions,
we can use the identity (after proper regularization)
\begin{equation}
   \int \frac{dk}{4\pi} \omega(k) e^{-ikx}=
   \omega^2\!\left(i\frac{\partial}{\partial x}\right)
   \int \frac{dk}{4\pi} \frac{e^{-ikx}}{\omega(k)}.
%   \int \frac{dk}{4\pi} \sqrt{m^2+a k^2+b k^4}e^{-ikx}=
%   \left(m^2-a\frac{\partial^2}{\partial x^2}+b\frac{\partial^4}{\partial x^4} \right)
%   \int \frac{dk}{4\pi} \frac{e^{-ikx}}{\sqrt{m^2+a k^2+b k^4}}.
\end{equation}
where $\omega^2(i\partial/\partial x)$ is a shorthand for $m^2-a\,{\partial^2}
/{\partial x^2}+b\,{\partial^4}/{\partial x^4}$ in the present case.
With this
\begin{equation}
   G_l(x)=
   G_\phi(x)\; \omega^2\!\left(i\frac{\partial}{\partial x}\right) G_\phi(x),
\end{equation}
where we have neglected the singular, delta-function term of Eq.\ (\ref{phiphi}).
This term can be evaluated directly for large $x$, knowing the asymptotic form
of $G_\phi$ in the different regimes. Using Eqs.\ (\ref{VBS}), (\ref{AFasy}) and
(\ref{ICasy}) we obtain
\begin{eqnarray}
   G_l(x)\sim \left\{
       \begin{array}{ll}
          \displaystyle{\frac{e^{-2v_-x}}{x^2}} &
                                \mbox{if $\theta<\theta_{\rm vbs}$} \\
          0^{}                  & \mbox{if $\theta=\theta_{\rm vbs}$} \\
          \displaystyle\frac{e^{-2vx}}{x^2} [c_1+c_2 \cos(2ux+\alpha)]
            & \mbox{if $\theta>\theta_{\rm vbs}$},
       \end{array}
   \right. 
\end{eqnarray}
where the constants $c_1,c_2$ and $\alpha$ can be expressed straightforwardly with
$a,b$ and $m$.  
It is interesting to remark that $G_l$ is exactly zero for any $x>0$ at the
disorder point, i.e., the 2-boson
processes do not contribute to the equal time correlation function there.
This can be easily verified by calculating $\omega^2 (i\partial/\partial x)
G_\phi(x)$ using Eq.\ (\ref{VBS}) and the fact that $D=0$ at the VBS point. Here
$\langle S^z_n S^z_0 \rangle$ only contains the $G_\phi$ term in full accordance
with the exact solution. For other values of $\theta$, $G_l$ decays
twice as rapidly as $G_\phi$.

It is also of interest to
see what predictions our simple field theory gives for the analytic
properties of the ground state energy density and the gap. Using the 
mode expansion in Eq.\ (\ref{mode}), the Hamiltonian can be written as
\begin{equation}
   H = \int_{-\Lambda}^{\Lambda} dk\; \omega(k) \left[
             d^{\dag}_k d^{}_k + {1\over 2} \right],
\end{equation}
where $\Lambda$ is an appropriate UV momentum cutoff, proportional to
the inverse of the lattice constant. From this the ground state energy is
\begin{equation}
   E = {1\over 2}\int_{-\Lambda}^{\Lambda} dk\; \omega(k).
\end{equation}
The ground state energy
depends analytically  on $a$, $b$ and $m$ whenever the zeros of $\omega(k)$
are not on the real axis. Together with the supposed analyticity of
$a(\theta)$, etc., this means that $E(\theta)$ is also analytic inside the
whole Haldane phase.\cite{Tit} A straightforward expansion around $\theta_{\rm vbs}$
yields
\begin{eqnarray}
   \lim_{\Lambda\to\infty} &&\left[ E(\theta)-E(\theta_{\rm vbs})\right] =\\
   && -\frac{8\pi a^{3/2}}{\sqrt{2} b}\sum_{n=1}^\infty
      \frac{2^{-6n}(4n-4)!}{(2n-2)!n!(n-1)!}  \left( \frac{D}{a^2}  \right)^n,
      \nonumber
\end{eqnarray}
which is convergent if $|D|/a^2<1$. We notice that
in general $E(\theta)$ can be expressed in a closed form in terms of
elliptic integrals of the first and second kind. 

The energy gap of the model is by definition
\begin{equation}
   \Delta = {\rm min}_k\; \omega(k) .
\end{equation}
This is obviously analytic at $\theta_{\rm cic}=\theta_{\rm vbs}$ but has a singularity
at $\theta_{\rm disp}$, where the minima of $\omega(k)$ move away from
$k=0$ as the parameter $a=0$ changes sign. While for
$\theta\le \theta_{\rm disp}$ the minimum is taken at $k=0$,
for $\theta>\theta_{\rm disp}$ it is taken at $k=\pm\sqrt{-a/2b}$. The gap
$\Delta$ and its derivatives with respect to $\theta$ on the two sides
of $\theta_{\rm disp}$ turn out ot be, resp.,
\begin{eqnarray}
   \Delta(\theta_{\rm disp}-0)&=&m, \quad\;\;\;\,
      \Delta(\theta_{\rm disp}+0)=m, \nonumber\\ \\
   \Delta' (\theta_{\rm disp}-0)&=&m', \quad\;\,
      \Delta' (\theta_{\rm disp}+0)=m',\nonumber\\
   \Delta'' (\theta_{\rm disp}-0)&=&m'', \quad
      \Delta'' (\theta_{\rm disp}+0)=
               m''-\left.\frac{\displaystyle{a'^2}}{\displaystyle{4bm}}
                          \right|_{\theta_{\rm disp}}. \nonumber
\end{eqnarray}                          
%\begin{equation}
%\begin{array}{ll}                          
%   \left. \Delta \right|_{\theta_{\rm disp}-0}=m,
%    & \left. \Delta \right|_{\theta_{\rm disp}+0}=m, \\ 
%   \left. \Delta' \right|_{\theta_{\rm disp}-0}=m', 
%    & \left. \Delta' \right|_{\theta_{\rm disp}+0}=m',\\
%   \left. \Delta'' \right|_{\theta_{\rm disp}-0}=m'',
%    & \left. \Delta'' \right|_{\theta_{\rm disp}+0}=
%               m''-\left.\frac{\displaystyle{a'^2}}{\displaystyle{4bm}}
%                          \right|_{\theta_{\rm disp}}.      
%\end{array}
%\end{equation}
We see that there is a discontinuity in the second derivative.
This behavior of the effective theory seems consistent with the
numerical results shown in Fig.\ 9 of Ref.\ [\ref{Sch-Jol-Gar}].

%%%%%%%%%%%%%%%%%%%%%%%%%%%%%%%%%%%%%%%%%%%%
\section{Effective theory - lattice version}

The effective theory presented above is capable of providing a
complete qualitative description of the C-IC transition
of the spin-1 bilinear-biquadratic model in accordance with the
available numerical data. However, in order to give quantitative
predictions, too, the theory needs some refinement. We have seen
earlier that our simple theory identifies the disorder point
$\theta_{\rm cic}$ with the point where the discriminant $D$ defined
by Eq.\ (\ref{discriminant}) vanishes. It is easy to verify that
the second and fourth derivatives of $\omega(k)$ at $k=0$ are, resp.,
\begin{equation}
   \frac{\partial^2\omega}{\partial k^2}(k=0)={a\over m},\qquad
   \frac{\partial^4\omega}{\partial k^4}(k=0)=-{3D\over m^3},
   \label{derivs}
\end{equation}
thus in the above theory the fourth derivative vanishes at the disorder
point. In contrast with this, Golinelli et al.\cite{Gol-Jol-Sor}
measured numerically the second and fourth derivatives at the known
disorder point $\theta_{\rm vbs}$ and found
\begin{equation} 
   \frac{\partial^2\omega}{\partial k^2}(k=0)=0.9778(1),\quad
   \frac{\partial^4\omega}{\partial k^4}(k=0)=-1.202(1).
   \label{numVBS}                     
\end{equation}
Note that the fourth derivative is only zero 
far inside the IC regime, which seems inconsistent with the above
theory. 

One step to improve the theory is to realize that the model is
defined on a lattice, and thus the dispersion $\omega(k)$
must be a $2\pi$-periodic function of $k$ (from now on the lattice
constant is set $\delta x=1$). This can be incorporated into the
Lagrangian in Eq.\ (\ref{Lagr}) by the standard
replacement $\partial_x\phi \to
[\phi(n+\delta x)-\phi(n)]/\delta x$, leading to the substitution
$k^2\to 2[1-\cos(k)]$ in the dispersion
\begin{eqnarray}
   \omega (k)&& = \\
   && \sqrt{m^2+2a+6b - (2a+8b)\cos(k) + 2b\cos(2k)}.    \nonumber 
\end{eqnarray}
The Green's function now reads
\begin{equation}
   G_\phi(n) =
   \int_{-\pi}^\pi \frac{dk}{4\pi} \frac{1}{\omega(k)}e^{ikn}. 
   \label{G_discrete}
\end{equation}
The condition that the expression under the square root in $\omega(k)$ is a
complete square is again $D(\theta)=0$ with $D$ defined in Eq.\
(\ref{discriminant}). When this is satisfied
the dispersion simplifies to
\begin{equation}
   \omega_{\rm vbs}(k)=
      m_{\rm vbs}+\frac{a_{\rm vbs}}{m_{\rm vbs}}[1-\cos(k)],
   \label{dispCIC2}
\end{equation}
and $1/\omega_{\rm vbs}(k)$ has poles instead of branch cuts. [One pole
within the Brillouin zone $-\pi<{\rm Re}(k)\le \pi$ with
$0<{\rm Im}(k)$.] Now the second and fourth derivatives at $k=0$ are
[cf.\ Eq.\ (\ref{derivs})]
\begin{equation}
   \frac{\partial^2\omega}{\partial k^2}(k=0)={a\over m},\qquad
   \frac{\partial^4\omega}{\partial k^4}(k=0)=-{3D\over m^3}-{a\over m}.
\end{equation}
At the disorder point we find
\begin{equation}
  \frac{\partial^4\omega}{\partial k^4}(k=0)=-{a_{\rm vbs}\over
  m_{\rm vbs}},
  \label{om4}
\end{equation}
which is nonzero. The correlation length $\xi=1/{\rm Im}(k)$
at the disordered point is determined by the position of the pole,
i.e., by the solution of the transcendental equation
\begin{equation}
   1+\frac{a_{\rm vbs}}{m^2_{\rm vbs}}[1-\cos(k)] = 0.
   \label{tranc}
\end{equation}
Working the other way around, knowing that at the VBS (C-IC) point
$\xi_{\rm vbs}=1/\ln 3$, Eq.\ (\ref{tranc}) gives
$a_{\rm vbs}=3m^2_{\rm vbs}/2$ and thus the dispersion
\begin{equation}
   \omega_{\rm vbs} = m_{\rm vbs}\left[{5\over 2}-{3\over 2}\cos(k)
                                 \right].
   \label{singlemode}                              
\end{equation}
With this expression the lattice Green's function defined in Eq.\ (\ref{G_discrete}) reads
\begin{equation}
  G_\phi(n) = {1\over 4m_{\rm vbs}} e^{-|n|\ln 3},
  \label{lG}
\end{equation}
which should be conferred to Eq.\ (\ref{G_VBS}).
The functional form of the dispersion relation in Eq.\ (\ref{singlemode})
is exactly the same
as the one appearing in the {\em single mode approximation} of the VBS
model.\cite{Fat-Sol-CM,Aro-Aue-Hal} There, one derives an upper bound
for the gap $\Delta_{\rm vbs}=m_{\rm vbs}=\sqrt{40}/9$, whereas here
we should use the phenomenological (numerical) value
$\Delta_{\rm vbs}=0.664314$ in Eq.\ (\ref{singlemode}).
This, together with Eq.\ (\ref{om4}) yields the value
$\partial^4\omega/\partial k^4=-3m_{\rm vbs}/2 \approx -1.0$,  %-0.0415$,
which is rather close to the numerical estimate in
Eq.\ (\ref{numVBS}).

The split of the double root in the vicinity of the disorder
point can be analyzed similarly to the continuum theory. We do not go
into details here, but emphasize that the critical exponents
characterizing the behavior of $\xi(\theta)$ and $q(\theta)$ at
$\theta_{\rm vbs}$, and the type of the singularity of the gap
at $\theta_{\rm disp}$ remain the same.
Similarly, we find that the ground state energy is analytic everywhere.

One can wonder about the possible consequences of keeping higher order
terms in the continuum Lagrangian Eq.\ (\ref{Lagr}) or in the
improved lattice
version. If the theory remains free the only effect is to bring about
additional branch cuts or poles in $1/\omega(k)$. If the higher order
terms are small the additional branching points are far in $k$ space, and
the C-IC transition remains intact. The long distance
asymptotics of the correlation functions do not change.
Since at the VBS point the exact correlation function in
Eq.\ (\ref{G_VBS}) only contains a single exponential term, the free
boson approximation does not allow any higher order spatial derivatives
in the effective Lagrangian there.

Although the dispersion $\omega_{\rm vbs}$ in Eq.\ (\ref{singlemode}) 
gives rise to the exponential term in Eq.\ (\ref{G_VBS}), it misses the
$\delta_{n,0}$ contribution. This is, however, another artifact of the
continuum approach, which necessarily neglects some important
short distance details. In fact, we should recall that in a quantum
spin liquid with short-range valence bond ground state the elementary
excitations (bosons) are physically triplet bonds living {\em between} lattice
sites, rather than on the sites themselves.
As was argued in Ref.\ [\ref{Fat-Sol-CM}], $S^z_i$ acting on the
VBS ground state produces the linear combination of two states, one
of which containing a boson at site $i-1/2$, the other a boson at site
$i+1/2$. Hence the one-boson term of $\langle S_i S_{i+n} \rangle$ is
in fact
\begin{equation}
   \langle S_n^z S_0^z\rangle = g_\phi^2 (-)^{n} [G_\phi(n-1)+
                   2G_\phi(n)+G_\phi(n+1)].
   \label{GGGG}
\end{equation}
Using Eq.\ (\ref{lG}) this leads directly to Eq.\ (\ref{G_VBS}), including the
$\delta$-function piece, if $g_\phi^2=m_{\rm vbs}$.

\section{Summary and discussion}

In summary, we proposed a simple effective field theory to describe
the commensurate-incommensurate
transition in the Haldane phase of the spin-1 bilinear-biquadratic chain. The theory is
capable of reproducing many features of this transition previously seen in the numerical
studies. Moreover it also has some new predictions.
The effective theory predicts that {\it the C-IC transition at $\theta_{\rm vbs}$
is not a phase transition in the
conventional sense}, since the ground state energy remains an analytic function of the
control parameter $\theta$. The only singularity occurring is in the correlation length.
We should emphasize, however, that the correlation function itself remains analytic as
a function of $\theta$ for any fixed distance, unlike in conventional phase
transitions.

There is another point $\theta_{\rm disp}$ close to the disorder point where another
quantity becomes singular. This is the energy gap (Haldane gap) whose second
derivative produces a jump. This singularity in the singlet-triplet gap becomes important
when a high enough magnetic field is applied, producing a crossing between these levels,
and thus leading to the collapse of the gap. At the critical field, as $\theta$ is varied
a real
phase transition takes place at $\theta_{\rm disp}$, thus this point is the endpoint
of a phase transition line on the magnetic field vs $\theta$ plain separating
two Luttinger liquid type phases.

In a technical sense the C-IC transition is a consequence of an accidental degeneracy
of roots of the dispersion relation. We have shown in the free boson approach
that this degeneracy causes the
"dimensional reduction" of the correlation function, and makes it to be a pure
exponential at the VBS point. We also found that the two-magnon contributions, and
presumably any higher order, multi-magnon contributions
too, vanish exactly at this point. We derived a formula valid in the vicinity of the
VBS point showing how the pure exponential decay emerges from the standard form with
algebraic prefactors. In particular we found that there is a crossover between a pure
exponential decay and a decay containing algebraic prefactors. The characteristic
distance of this crossover tends to infinity as the VBS point is approached.

The spin-1 bilinear-biquadratic model studied in this paper is not the only model
which produces a C-IC transition. Another example is the spin-1/2 chain
with next-nearest-neighbor interaction (this can also be visualized as a two-leg
zig-zag ladder).\cite{watanabe}
By now it is well established that the Majumdar-Ghosh point of
this model is a disorder point where a C-IC transition of the first kind occurs.
On the same footing as described here,
it seems possible to develop an effective theory which supposes
that the elementary excitations (spin-1/2 solitons in that case) are essentially free
particles with a non-relativistic dispersion. Care should, however, be taken on the facts
that solitons are always created in pairs and that they are spin-1/2 particles.
Beside the $S=1/2$ case, the appearence of disorder points has been demonstrated
in other $S\ge 1$ frustrated Heisenberg chains too.\cite{KoleRoth} Another interesting
quasi-one-dimensional system where a commensurate-incommensurate
transition have been reported in a numerical investigation is the SU(2)$\times$SU(2)
symmetric coupled spin-orbit model.\cite{Patietal}
The elaboration and testing of effective theories,
similar to the one described in this paper, for these models could be
a possible direction of future research.

We thank L. Balents and J. S\'olyom for valuable discussions. This work was financially
supported by the Hungarian Scientific Research Found (OTKA) under grant Nos. 30173,
30543, and F31949.

\end{document}